\documentclass{aa}
\usepackage{graphicx}
\usepackage{amsmath}

\newcommand{\kms}{\mbox{km\thinspace s$^{-1}$}}

\def\to {$\rightarrow$}
\def\CI{[C{\sc i}]}
\def\CII{[C{\sc ii}]}

\begin{document}

\title{Atomic carbon in PSS~2322+1944, a quasar at redshift $4.12$
  \thanks{Based on observations obtained with the IRAM Plateau de Bure
    interferometer.}}

\author{J. Pety\inst{1,2},
        A. Beelen\inst{3},
        P. Cox\inst{3},
        D. Downes\inst{1},
        A. Omont\inst{4},        
        F. Bertoldi\inst{5},                
        C.L. Carilli\inst{6}         
        }

\offprints{J. Pety, pety@iram.fr}

\institute{IRAM, 300 rue de la Piscine, 38406 St-Martin-d'H\`eres, France
  \and LERMA, Observatoire de Paris, F-75014 Paris, France
  \and Institut d'Astrophysique Spatiale, Universit\'e de Paris Sud, F-91405 Orsay, France
  \and Institut d'Astrophysique de Paris, CNRS and Universit\'e de Paris VI, 
       98b bd. Arago, F-75014 Paris, France 
  \and Max-Planck-Institut f\"ur Radioastronomie, Auf dem H\"ugel 69, D-53121 Bonn, Germany
  \and National Radio Astronomy Observatory, P.O. Box, Socorro, NM~87801, USA}

\date{Received September 21, 2004 / Accepted October 27, 2004}

\titlerunning{\CI{} in a quasar at redshift 4.12}
\authorrunning{J. Pety et al.}

\abstract{We report the detection of the $^3P_1$\to$^3P_0$ fine-structure
  line of neutral carbon in the $z=4.12$ quasar PSS~2322+1944, obtained at
  the IRAM Plateau de Bure interferometer.  The \CI{} $^3P_1-^3P_0$ line is
  detected with a signal-to-noise ratio of $\sim 6$ with a peak intensity
  of $\rm \approx 2.5 \, mJy$ and a velocity-integrated line flux of $\rm
  0.81 \pm 0.12 \, Jy \, km \, s^{-1}$.  Assuming an excitation temperature
  of 43~K (equal to the dust temperature), we derive a mass of neutral
  carbon (corrected for magnification) of $M_{\rm C \sc I} \approx 1.2
  \times 10^7 \, \rm M_\odot$.  In PSS~2322+1944, the cooling due to C is
  about 6 times smaller than for CO, whereas the CO and C cooling
  represents $\approx 10^{-4}$ of the far-infrared continuum and more than
  half of the cooling
  due to C$^+$. %
  } %

\maketitle \sloppy

\section{Introduction}
\label{sec:Introduction}

With the detection of dust and molecular gas in sources at high redshift,
it has become possible to probe the physical conditions of the interstellar
medium in galaxies and in the hosts of quasi-stellar objects (QSOs) at
cosmological distances. The high-$z$ sources detected in $\rm ^{12}CO$ (to
date 30 sources between $1.44 < z < 6.42$ - see, e.g., Greve et al.  2004)
have massive reservoirs of warm and dense molecular gas (a few $\rm 10^{10}
- 10^{11} \, M_\odot$), which are predominantly excited by extreme
starbursts with implied star formation rates $\rm \approx 10^3 \, M_\odot\,
yr^{-1}$.  Multiline CO studies are available in only a few cases, and the
detection of species other than carbon monoxide is reported in only a
couple of sources. A remarkable example is the Cloverleaf, a strong
gravitationally lensed QSO at $z=2.56$ where four CO transitions were
detected, together with the two fine-structure lines of neutral carbon
\CI{} (Barvainis et al. 1997; Wei\ss~ et al.  2003) and the J=1\to0
transition of HCN (Solomon et al.  2003).

Atomic carbon is an important probe of the neutral dense gas. It is a good
tracer of molecular gas in external galaxies and plays a central role in
the cooling of the gas (G\'erin \& Phillips 1998, 2000).  In high-$z$
sources, the detection of the \CI{} lines enables to obtain further
constraints on the physical conditions of the interstellar gas in addition
to those obtained from the CO transitions (Wei\ss\ et al.  2003), providing
useful information on the gas column density, the thermal balance, and the
UV illumination.

PSS~2322+1944 is an optically luminous, gravitationally lensed QSO at
$z=4.12$ which was studied in detail both in the dust and radio continuum
emission (Omont et al.  2001; Beelen et al. 2004) and in the J=5\to4,
4\to3, 2\to1, and 1\to0 transitions lines of CO (Cox et al.  2002; Carilli
et al. 2002).  With an apparent far-infrared (FIR) luminosity of $\rm 3
\times 10^{13} \, L_\odot$, it harbors a massive reservoir of molecular gas
which is the site of active star formation with an implied star formation
rate of $\rm \sim 900 \, M_\odot \, yr^{-1}$. The CO line emission is
resolved into an Einstein Ring with a diameter of $1\farcs5$, a direct
indication of lensing.  The amplification factor is estimated to be about
3.5. The data are consistent with a disk surrounding the QSO with a radius
of 2~kpc and a dynamical mass of a few $\rm 10^{10} \, M_\odot$ (Carilli et
al.  2003).

\begin{figure}
  \includegraphics[width=\hsize]{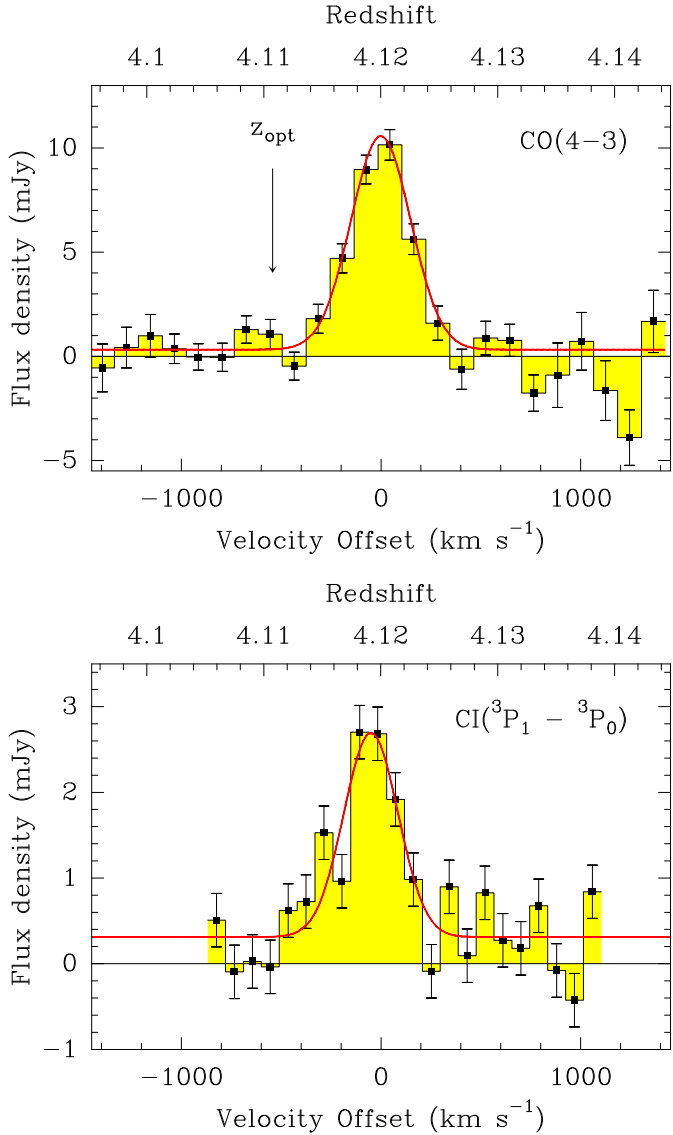}
  \caption{Observed spectra of the \CI{}$^3P_1$\to$^3P_0$ (this paper)
    and CO($\rm J=$4\to3) lines (from Cox et al. 2002) toward the $z=4.12$
    quasar PSS~2322+1944. The black horizontal lines show the zero flux
    density level. The red lines show the results of Gaussian + continuum
    fits.}
\label{fig:spectra}
\end{figure} 

\begin{figure}
  \includegraphics[height=\hsize,angle=270]{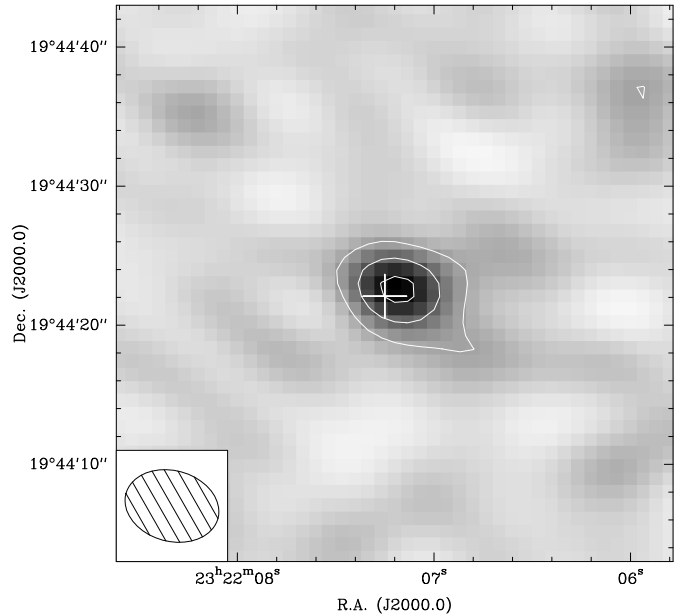}
  \caption{Map of the \CI{} $^3P_1$\to$^3P_0$ velocity-integrated emission 
    toward PSS~2322+1944. The emission is integrated between $-400$ and
    $+400~\kms$. The white cross indicates the optical position at
    R.A.=23:22:07.25, Dec.=19:44:22.08 (J2000.0). The contours correspond
    to multiples of $3\sigma = 0.34$~mJy/beam. Negative levels are shown as
    dotted contours with the same step.  The synthesized beam of $\rm
    6\farcs8 \times 5\farcs1$ ($73\degr$) is shown in the lower left
    corner.  }
\label{fig:map} 
\end{figure} 

The observations of the molecular gas in PSS~2322+1944 provide one of the
best examples to date of active star formation in the host galaxy of a
luminous, high redshift QSO.  The strong CO emission in PSS~2322+1944 makes
it a prime target to search for the $^3P_1$\to$^3P_0$ transition of \CI{}
($\nu_\mathrm{rest} = 492.161$~GHz) which, at $z=4.12$, is shifted into the
3~mm atmospheric window.

Here we report the $^3P_1$\to$^3P_0$ \CI{} line in PSS~2322+1944.  After
the detection of \CI{} in the Cloverleaf, and, recently, in IRAS FSC 10214
($z=2.3$) and SMM~J14011+0252 ($z=2.5$) reported by Wei\ss~ et al. (2004),
these observations represent the fourth clear detection of \CI{} in a
high-$z$ source. In this paper, we assume the concordance
$\Lambda$-cosmology with $H_0=71\rm~km~s^{-1}~Mpc^{-1}$,
$\Omega_\Lambda=0.73$ and $\Omega_m=0.27$ (Spergel et al. 2003).

\begin{table*}[ht]
  \caption{Properties of the \CI{} and CO(4\to3) lines observed toward
    PSS~2322+1944.}
  \begin{center}
    \begin{tabular}{ccccccccc}
      \hline
      \hline 
      Line &  $\rm \nu_{\rm rest}$ & $\rm \nu_{\rm obs}$ & Peak Int. & Position & $\Delta v_{\rm FWHM}$ &    $I$      & $       L^\prime$            &       $L$ \\
           &          \multicolumn{2}{c}{[GHz]}          &   [mJy]   &   \multicolumn{2}{c}{[\kms{}]}   & [Jy~\kms{}] & [$\rm 10^{10} K \, \kms \, pc^2$] & [$\rm 10^8 \, L_\odot$] \\ 
      \hline
      \CI{} $^3P_1$\to$^3P_0$  & 492.161  & 96.127  &  2.4 & $-$52$\pm$25 & 319$\pm$66 & 0.81$\pm$0.12 & $ 3.4 \pm 0.5$ & $1.1 \pm 0.2$ \\     
      CO (4\to3)$^\dagger$     & 461.041  & 90.048  & 10.3 &  $-$3$\pm$10 & 348$\pm$40 & 3.83$\pm$0.40 & $16.3 \pm 1.5$ & $5.1 \pm 0.5$ \\      
      \hline
    \end{tabular}
  \end{center}
  NOTE. -- The CO(4\to3) data are from Cox et al. (2002). The continuum and
  the line were fitted separately using an \emph{a priori} line window 
  (between $-400$ to $+400~\kms$) defined from the high signal-to-noise CO
  $J=$4\to3 spectrum. The luminosities are not corrected for the lens 
  amplification ($\rm m=3.5$).  
  \label{tab:properties}
\end{table*}

\section{Observations and data reduction}
\label{sec:observations}

Observations of the $^3P_1$\to$^3P_0$ transition of \CI{} in PSS~2322+1944
were carried out with the IRAM Plateau de Bure interferometer (PdBI) in a
series of observing sessions between July 2002 and June 2004.  The total
integration time of the useful data is equivalent to 34 hours with 6
antennas. In practice, all the observing time was allocated during the
summer, when antenna maintenance takes place. The total on-sky integration
time amounts to 116 hours often with 4 or 5 (and sometimes with 6)
antennas.

We used the interferometer in the D configuration.  The combined 3.2~mm
data result in a synthesized beam of $6\farcs85 \times 5\farcs09$ at a
position angle of $73\degr$. The 3~mm receivers were tuned first at the
red-shifted frequency of the \CI{} $^3P_1$\to$^3P_0$ line, i.e. 96.127~GHz
at $z=4.119$ and then at 96.063~GHz for the second half of the data.  This
is the reason why the spectrum is noisier at the very ends of the band
(Fig.~\ref{fig:spectra}).  At those frequencies, the typical SSB system
temperatures were $\approx 150-250$~K. The water vapor ranged between 4 and
10~mm on different sessions. The 580~MHz instantaneous IF--bandwidth were
observed with a resolution of 1.5~MHz.

All data reduction were done using the \textsc{gildas} softwares.  Standard
calibration methods using close calibrators were applied. The bandpass
calibration was done on the quasar 3C454.3.  The amplitude and phase
calibration were performed on 3C454.3 and the nearby quasar 2230+114. Only
data with phase noise better than 40~deg were used. The maximum position
errors at 3.2~mm introduced by such phase noise is $< 0\farcs5$.  The flux
calibration is based on the PdBI primary calibrator MWC~349.  The fluxes of
3C454.3 (resp.  2230+114) varied from 7 to 4~Jy (resp. 4 to 2.4~Jy) during
the 3 years observing period.

A standard calibrated $uv$ table was produced and analyzed both by direct
fits in the $uv$ plane (as this avoids the deconvolution step) and by
making the deconvolved image shown in Fig.~\ref{fig:map}. Both methods
confirm that the \CI{} $^3P_1$\to$^3P_0$ emission is \emph{i)} detected,
\emph{ii)} not resolved by our observations and \emph{iii)} centered on the
optical position which is also the phase center of the observations.  The
two latter points allow us to derive the source spectrum from the real part
of the average of all the complex visibilities.  The main advantage of this
method is to directly obtain the spectrum from simple operations instead of
applying the usual fitting procedures in the $uv$ plane.  Any further
processing (smoothing, fitting) was performed using the \textsc{class}
software.

The final \CI{} spectrum at a velocity resolution of 90~\kms{} is displayed
in Fig.~\ref{fig:spectra} together with the spectrum of the $J=$4\to3
transition of CO.  The image of the \CI{} emission is shown in
Fig.~\ref{fig:map}. The \CI{} $^3P_1$\to$^3P_0$ fine-structure line is
clearly detected at the same redshift as the CO emission with a peak flux
density of $S_\nu = 2.69 \pm 0.31 \, \rm mJy$ (including the continuum flux
of 0.31~mJy).  Table~\ref{tab:properties} summarizes the line parameters
derived from continuum + Gaussian fits.

\section{Results and Discussion}
\label{sec:results}

The integrated flux of the \CI{} $^3P_1$\to$^3P_0$ line is $\rm 0.81 \pm
0.12 \, Jy \, km \, s^{-1}$, a factor of 5 lower than the CO(4\to3) line. A
Gaussian fit yields a line width of $\rm 319 \pm 66 \, km \, s^{-1}$ and a
line center displaced by $\rm -50 \, km \, s^{-1}$ relative to CO(4\to3).
Due to the weak signal to noise ratio of the \CI{} data, the difference in
the line position with the higher quality CO(4\to3) data should not be
over-interpreted.  The \CI{} line flux implies a line luminosity of
$L^\prime_\mathrm{CI} = 3.4 \pm 0.5 \times 10^{10} \, \rm K \, km \, s^{-1}
\, pc^2$ or $\rm 1.1 \pm 0.2 \times 10^8 \, L_\odot$ (see, e.g., Solomon et
al. 1997 for the definition of the line luminosity).  The \CI{} luminosity
in PSS~2322+1944 is thus about a factor 2 lower than in the case of the
Cloverleaf where $L^\prime_\mathrm{CI} = 6.1 \times 10^{10} \, \rm K \, km
\, s^{-1} \, pc^2$, as derived from the \CI{} velocity-integrated flux of
Barvainis et al.  (1997).

The continuum emission is detected at 96~GHz with $0.31 \pm 0.08 \, \rm
mJy$, which is consistent for dust emission with the available photometric
data at higher frequency. The FIR spectral energy distribution of
PSS~2322+1944 is well reproduced with a grey body of temperature $43\pm
6$~K and a dust emissivity $\propto \nu^{1.6\pm0.3}$ (Beelen et al. 2004).
   
The detection of the \CI{} $^3P_1$\to$^3P_0$ emission line in PSS~2322+1944
allows us to estimate the mass of neutral carbon $M_\mathrm{CI}$. Since the
$^3P_2$\to$^3P_1$ transition (at $\nu_\mathrm{rest} = 809.342$~GHz) is not
observed, we assume that the $^3P_1$\to$^3P_0$ transition is optically thin
as is the case for the Cloverleaf (Wei\ss\ et al.  2003). Under these
assumptions, we can express the mass of neutral carbon as a function of
$L^\prime_\mathrm{CI}$
\begin{equation}
  \label{eq:mci}
  M_\mathrm{CI} = 5.65 \times 10^{-4} \, {Q\over 3} \,
  \mathrm{e}^{(23.6/T_\mathrm{ex})} \,  L^\prime_\mathrm{CI (^3P_1\rightarrow^3P_0)}  
  \, \, \, \, \mathrm{M_\odot}, 
\end{equation}
where $T_\mathrm{ex}$ is the excitation temperature and $Q$ is the C{\sc i}
partition function. Assuming that $T_\mathrm{ex}$ is equal to the
temperature of the warm dust $T_\mathrm{dust} = 43~\mathrm{K}$, the mass of
neutral carbon in PSS~2322+1944 amounts to $M_\mathrm{CI} = 4.3 \times 10^7
\, \mathrm{M_\odot}$, or $1.2 \times 10^7 \, \mathrm{M_\odot}$ after
correction for lens amplification ($\rm m=3.5$).  Although the excitation
temperature of the neutral carbon can be different than $T_\mathrm{dust}$
(as observed for the Cloverleaf - see Wei\ss\ et al.  2003), the derived
mass depends only weakly on $T_\mathrm{ex}$. For a range of \CI{}
excitation temperatures from 30 to 100~K, the mass of neutral carbon in
PSS~2322+1944 would vary from 1.2 to $1.4 \times 10^7 \, \mathrm{M_\odot}$.

\begin{table*}
  \caption{Comparison of the interstellar gas and dust luminosities in Infrared Luminous Galaxies
    and in the Galactic Center}
  \begin{tabular}{cccccccccccc}
    \hline
    \hline \\[-0.2cm]
    Source         & $z$ & C$^\dagger$  & CO$^{\dagger\dagger}$ & C$^+$ & $\rm L_{FIR}$ & CO/C & (C+CO)/$\rm L_{FIR}$ & $\rm C^+/\rm L_{FIR}$ & Ref.  \\  
    & & [L$_\odot$] & [L$_\odot$] & [L$_\odot$] & [L$_\odot$] & & & & &           \\
    \hline \\[-0.2cm]
    PSS~2322+1944$^{(a)}$ & 4.12   & $1.0 \times 10^8$ & $6.8 \times 10^8$ & $\le 1.7 \times 10^9$ & $8.6 \times 10^{12}$   & 6.6  & $9 \times 10^{-5}$   & $\le 2 \times 10^{-4}$ & [1] \\ 
    Cloverleaf$^{(a)}$    & 2.56   & $7.1 \times 10^7$ & $1.5 \times 10^{9}$ &       --         & $4.2 \times 10^{12}$ & 21.7 & $3.9 \times 10^{-4}$ & --                     & [2] \\
    Arp220                & 0.018  & $1.8 \times 10^7$ & $3.9 \times 10^7$ &   $1.6 \times 10^9$ & $1.2 \times 10^{12}$ & 2.2  & $4.7 \times 10^{-5}$ &   $1.3 \times 10^{-3}$ & [3] \\
    NGC253                & 0.0008 & $1.6 \times 10^5$ & $1.6 \times 10^6$ &   $7.8 \times 10^6$ & $1.0 \times 10^{10}$ & 10.0  & $1.8 \times 10^{-4}$ &   $7.8 \times 10^{-4}$ & [4] \\
Galactic Center$^{(b)}$   &     -- & $6.7 \times 10^4$ & $2.6 \times 10^5$ &   $2.6 \times 10^6$ & $3.9 \times 10^8$    & 3.8  & $8.4 \times 10^{-4}$ &   $6.7 \times 10^{-3}$ & [5] \\
    \hline
  \end{tabular}
  NOTE. -- $^\dagger$ Total {\sc C i} luminosity. $^{\dagger\dagger}$ 
  CO luminosity up to $J=8$. $(a)$ The luminosities are 
  corrected for lensing: $\rm m=3.5$ for PSS~2322+1944 (Carilli et al. 2003)
  and $m=11$ for the Cloverleaf (Venturini \& Solomon (2003). $(b)$ The
  central $5 \times 1 \, \rm deg^2$ of the Galaxy. References: [1] This paper -- [2]
  Wei\ss~ et al. (2003); Barvainis et al.  (1997); A. Wei\ss (private
  communication) -- [3] G\'erin \& Phillips (1998) -- [4] Bayet et
  al. (2004) -- [5] Fixsen et al. (1999).
  \label{tab:luminosities}
\end{table*}

Compared to the mass of molecular gas of $M_\mathrm{H_2} = 7 \times 10^{10}
\, \mathrm{M_\odot}$ after correcting for amplification (Cox et al.  2002;
Carilli et al. 2003), the derived mass of \CI{} implies a carbon abundance
relative to H$_2$ of \CI{}/[H$_2$]$\approx 3 \times 10^{-5}$, indicating
near to solar abundances in this high-redshift system. Similar values are
derived by Wei\ss~ et al. (2004). This relative carbon abundance is close
to the maximum value of $2.2 \times 10^{-5}$ found for Galactic dense
molecular clouds with opacities of $4 - 11 \, \mathrm{mag}$, a value which
does not vary within a factor of a few for larger $A_V$'s (Frerking et al.
1989).

The \CI{} $^3P_1\rightarrow^3P_0$ and $^3P_2$\to$^3P_1$ lines are major gas
coolants. In the Cloverleaf, the $^3P_2$\to$^3P_1$ line is 2.2 times
stronger than the $^3P_1\rightarrow^3P_0$ line (Wei\ss~ et al. 2003).  To
estimate the carbon cooling, we may assume a similar ratio for
PSS~2322+1944.  To compare this to the CO cooling, we added the observed CO
line luminosities of PSS~2322+1944 up to $J=9-8$, where for the unobserved
transitions we adopt the prediction of the LVG model of Carilli et al.
(2002). In PSS~2322+1944, we find that the CO/\CI{} luminosity ratio is 6,
as compared to a ratio of 20 for the Cloverleaf (see
Table~\ref{tab:luminosities}). However, the total mass of \CI{} remains
somewhat smaller than that of CO (see Carilli et al.  2002).

Compared to the far-IR luminosity of PSS~2322+1944, $L_\mathrm{FIR} \sim
8.6 \times 10^{12} \, \mathrm{L_\odot}$ (corrected for lensing), the CO and
\CI{} cooling represents $\sim 10^{-4}$ of the far-IR continuum, again not
very different from the ratio of $4 \times 10^{-4}$ derived for the
Cloverleaf (Table~\ref{tab:luminosities}). The recent search of the
red-shifted \CII{} fine-structure line in PSS~2322+1944 implies an upper
limit to the \CII{} line luminosity of $1.7 \times 10^9 \,
\mathrm{L_\odot}$, a weakness which is typical for high-$z$ IR luminous
galaxies (Benford et al. 2004). The CO and C cooling is therefore more than
half of the cooling due to $\rm C^+$. The \CII{} line remains the main
cooling line of the gas.
  
Finally, the \CI{} luminosity relative to the integrated far-IR luminosity
is a good measure of the intensity of the non-ionizing UV radiation field
in galaxies, because in photodissociation regions the column density of
neutral carbon is mostly insensitive to the UV field, whereas the far-IR
emission is directly proportional to the strength of the UV field (see,
e.g., Kaufman et al. 1999; G\'erin \& Phillips 2000). For PSS~2322+1944,
this ratio is $\sim 3 \times 10^{-6}$ indicating a UV radiation field of a
few 1000 times larger than in the solar vicinity. Both this ratio and the
implied strength of the UV illumination are comparable to the values
derived for the other IR luminous galaxies and the galactic center listed
in Table~\ref{tab:luminosities}.

As in the case for local starburst galaxies such as NGC253, CO is a more
important coolant than C in both PSS~2322+1944 and the Cloverleaf by about
one order of magnitude (Table~\ref{tab:luminosities}) - see also Schilke et
al. (1993) and Bayet et al. (2004).  Similarly, in starburst galaxies or in
galactic nuclei (including the Milky Way), the molecular gas is warm
therefore populating the higher CO levels which contribute to the cooling.

The detection of the red-shifted \CI{} $^3P_1$\to$^3P_0$ transition line in
the gravitationally lensed $z=4.12$ QSO PSS~2322+1944 enables to further
constrain the physical conditions of the neutral gas and to compare the
major line (C, CO and $\rm C^+$) and far-IR luminosities in this high-$z$
galaxy. Together with the recent \CI{} detections in other sources at high
redshift (Wei\ss~ et al. 2004), these results illustrate the potential of
studying neutral carbon or species other than CO in high-$z$ sources, a
field which will clearly fully develop as soon as more sensitive
submillimeter arrays, such as the Atacama Large Millimeter Array (ALMA),
will become operational.

\acknowledgements

We thank the IRAM Plateau de Bure staff for their support in the
observations. IRAM is supported by INSU/CNRS (France), MPG (Germany), and
IGN (Spain). Estelle Bayet and the referee, Dr. Phil M. Solomon, are kindly
acknowledged for helpful comments.

\end{document}